\documentclass[%
twocolumn,
 amsmath,amssymb,
 aps,
prb,
]{revtex4}
\usepackage[dvipdfmx]{graphicx}
\usepackage{color}  
\usepackage{comment}
\usepackage{dcolumn}
\usepackage{bm}

\begin{document}
\preprint{APS/123-QED}

\title{Bulk Rashba Effect in Multiferroics: a theoretical prediction for BiCoO$_{3}$}

\author{Kunihiko Yamauchi}
\affiliation{%
ISIR-SANKEN, Osaka University, 8-1 Mihogaoka, Ibaraki, Osaka, 567-0047, Japan
}%
\author{Paolo Barone}%
\author{Silvia Picozzi}
\affiliation{
Consiglio Nazionale delle Ricerche (CNR-SPIN), Sede Temporanea di Chieti, c/o Univ. ``G. D'Annunzio'' 66100 Chieti, Italy
}%

\date{\today}
\newcommand{\ba}{Ba$_{2}$CoGe$_{2}$O$_{7}$}
\newcommand{\bco}{BiCoO$_{3}$}

\begin{abstract}

We put forward the concept of a bulk Rashba effect emerging in a multiferroic material, such as an antiferromagnetic system with a polar crystal structure. According to symmetry considerations, while time-reversal and space-inversion symmetries are both broken, there exist specific spin flipping operations that relate opposite spin sites in the magnetic crystal structure. As a consequence, at certain high-symmetry points in the momentum space, the magnetic point group allows the spin angular momentum to be locked to the linear momentum, a typical feature of the Rashba effect. In such a case, spin-splitting effects induced by spin-orbit coupling can arise, similar to what happens in non-magnetic Rashba systems. As a prototypical example, ab-initio calculations of antiferromagnetic BiCoO$_{3}$ in the polar structure reveal that a large Rashba-like band- and spin- splitting occurs at the conduction band bottom, having a large weight from Bi-$p$ orbital states. Moreover, we show that the spin texture of such a multiferroic can be modulated by applying a magnetic field. In particular, an external in-plane magnetic field is predicted not only to induce spin-canting, but also a distortion of the energy isocontours and a shift of the spin-vortex (centered on the high-symmetry point and characteristic of Rashba effect) along a direction perpendicular to the applied field. 
\color{black}
\end{abstract}

\pacs{Valid PACS appear here}
\maketitle


\section{\label{sec:intro}Introduction}

\begin{figure}[ht!]
\begin{center}
{
\includegraphics[width=55mm, angle=0]{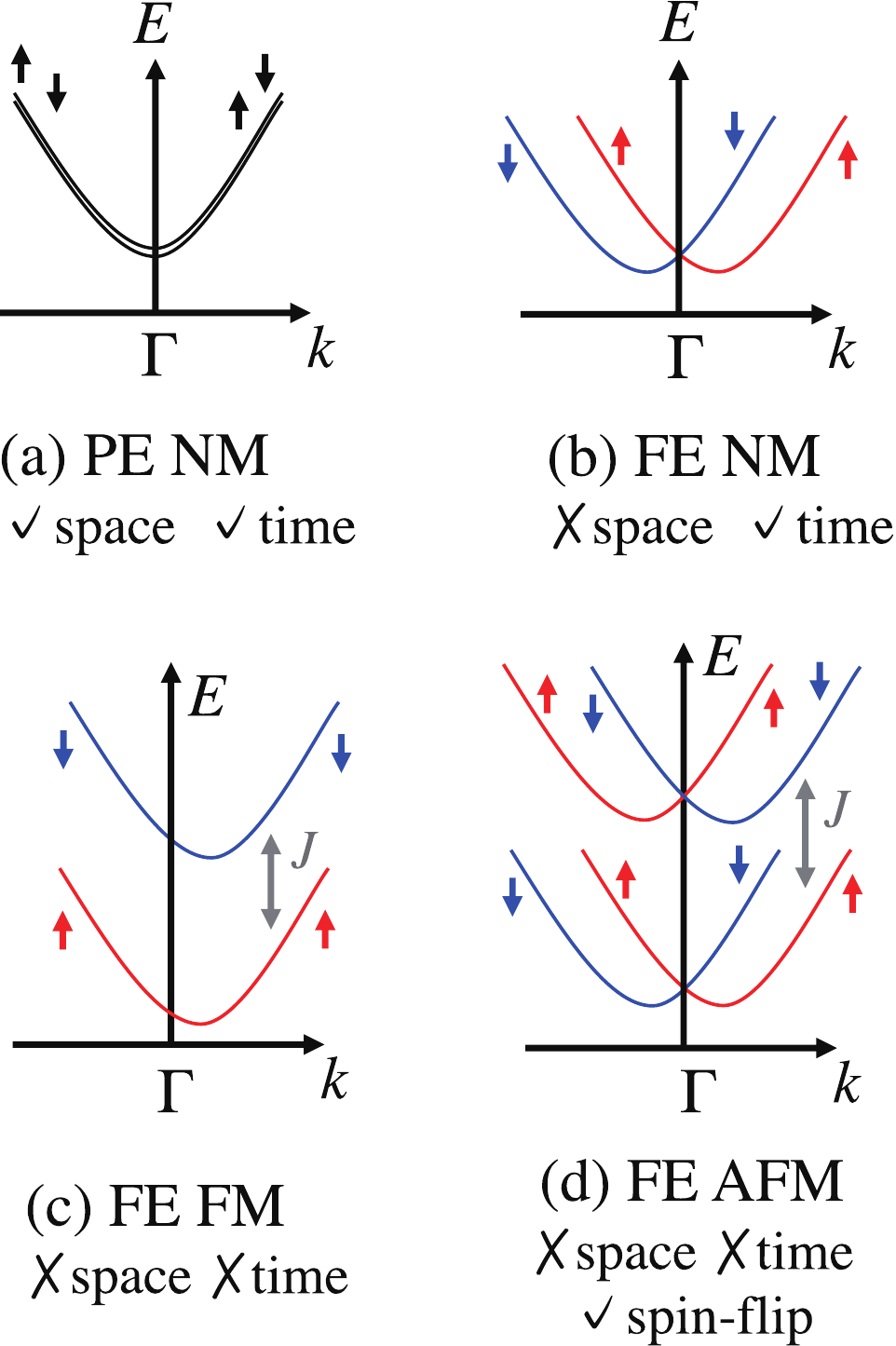}
}
\caption{\label{fig:pontier}
Schematic pictures of spin-split bandstructure for systems which are
(a) paraelectric and nonmagnetic, (b) ferroelectric and nonmagnetic, (c) ferroelectric and ferromagnetic, and (d) ferroelectric and antiferromagnetic. 
Bulk Rashba effect can take place in (b) and (d) cases. 
 Here, ``space'',  ``time'', and ``spin-flip'' denote space-inversion, time-reversal, and spin-flipping symmetry, respectively.
}
\end{center}
\end{figure}

Transition-metal oxides are widely used in spintronics, owing to the rich physical properties which often couple to various structural distortions. In case of non-magnetic materials with strong spin-orbit coupling (SOC), a polar ionic distortion can also remove the spin degeneracy of the bandstructure, with appealing potential for spin-FET applications.\cite{silvia.review} While spin-momentum locking phenomena, such as Rashba effect, 
commonly occur at interfaces or surfaces, 
it is nowadays well established that Rashba-like band splitting may also occur in non-centrosymmetric (chiral, polar or ferroelectric) bulk materials.\cite{BiTeI,domenico.gete,gete_exp1,gete_exp2,rashba1, rashba2, rashba3, rashba4,yamauchi.prl.biiro3,narayan,hfo2,varignon2019} 
%
Here we are taking a step further by considering a noncentrosymmetric antiferromagnetic oxide as a playground for bulk Rashba effect. 
The latter has been indeed observed in polar BiTeI\cite{BiTeI} and ferroelectric GeTe\cite{gete_exp1,gete_exp2}, as well as predicted in several ferroelectric materials.\cite{rashba1, rashba2, rashba3, rashba4,luiz.bialo3,yamauchi.prl.biiro3,narayan,hfo2,varignon2019} 
As schematically illustrated in Fig. \ref{fig:pontier}, bulk Rashba effect can occur  in ferroelectric nonmagnetic systems 
where the time-reversal symmetry relate up- and down-spin bands at $\bm k$ and $-{\bm k}$ points. 
In systems which are both ferroelectric and ferromagnetic, conversely, on-site exchange interaction $J$ largely splits the bands so that up- and down-spin bands become unpaired. 
In ferroelectric and antiferromagnetic systems, an alternative scenario occurs: spin-flipping symmetry operations may exist (such as a mirror symmetry operation) which behave as time-reversal symmetry, thus enforcing two-fold degeneracy at specific high-symmetric points. In this case, bulk Rashba effect can emerge.\cite{hotta2019}
 This possibility ultimately relies on the fact that time-reversal symmetry, which is always broken in a ferromagnet, can still form a symmetry element of the relevant magnetic space group in antiferromagnets in suitable combinations with spatial symmetry operations.
Following this idea, in this paper, the electronic structure, the spin texture, and its correlation with spin canting in a noncentrosymmetric antiferromagnetic oxide will be discussed. 

\begin{figure}[ht!]
\begin{center}
{
\includegraphics[width=75mm, angle=0]{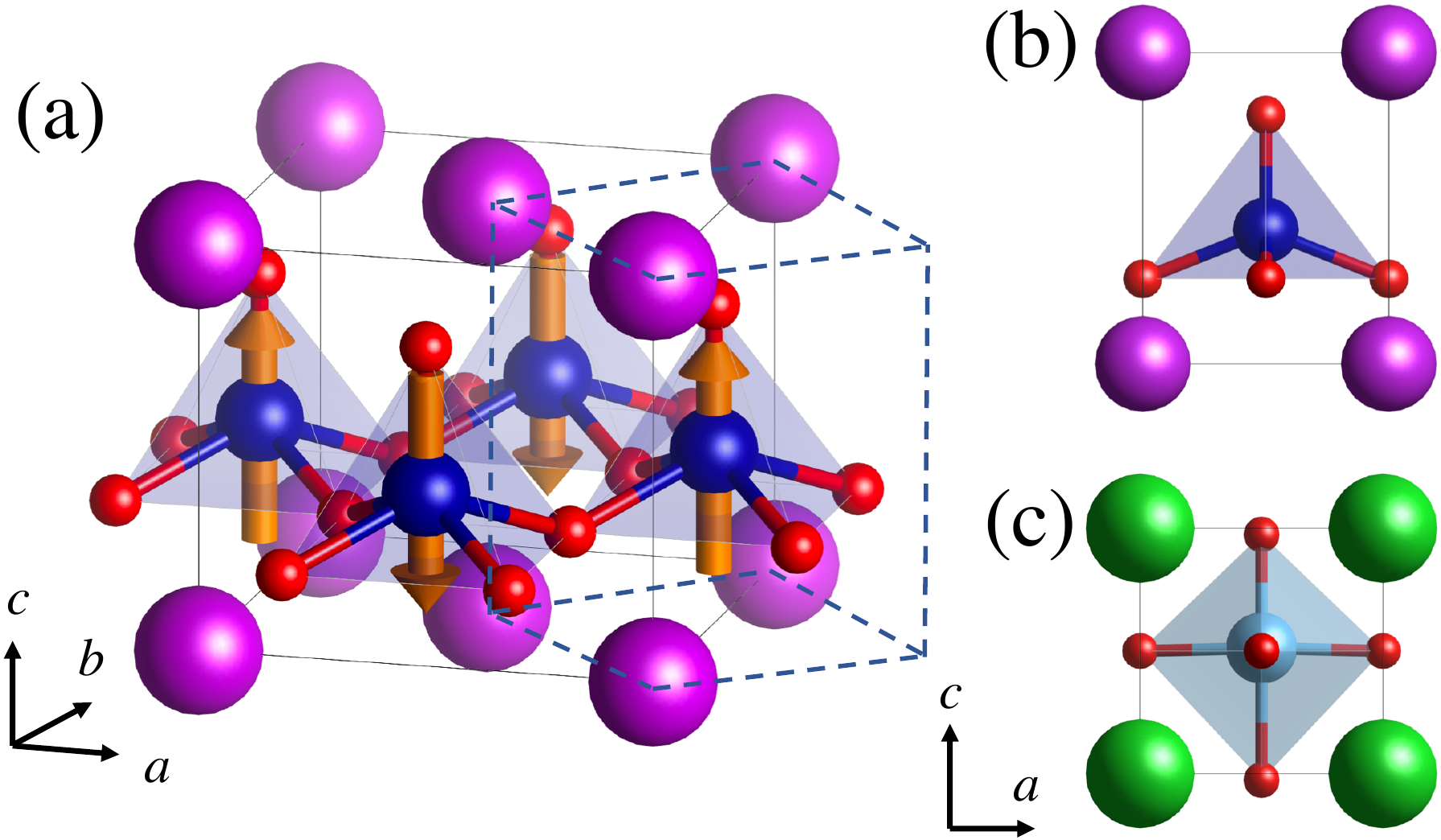}
}
\caption{\label{fig:crys}
(a) BiCoO$_{3}$ crystal structure with $P4mm$ space group. Magenta, blue, red spheres denote Bi, Co, O ions respectively. 
The nonmagnetic unit cell is shown by dashed lines whereas
we considered a $\sqrt{2}\times\sqrt{2}\times1$ super cell to host C-AFM spin ordering, as shown by orange arrows.    
Comparison of structure between (b) \bco\ and (c) BaTiO$_{3}$ in sideview, showing the pyramidal and octahedral coordination, respectively.
}
\end{center}
\end{figure}

Aiming at a sizable bulk Rashba spin-splitting in a magnetic system, here we propose 
BiCoO$_{3}$ as a candidate material. 
The perovskite BiCoO$_{3}$ was synthesized at high pressure of 6 GPa.\cite{belik.bicoo3} 
The crystal structure displays the polar $P4mm$ space-group symmetry; while the latter is often found in popular ferroelectrics (such as BaTiO$_{3}$ and PbTiO$_{3}$), 
\bco\ shows more prominent polar structural distortions. 
In the tetragonal lattice,  the Co$^{3+}$ ($d^{6}$) ion located at the center of the O$_{6}$ octahedron is significantly displaced toward the apical O ion, 
 leading to a pyramidal coordination rather than an octahedral coordination (see Fig. \ref{fig:crys}).
This large distortion is responsible for the high $c/a$ ratio of 1.267 and for the calculated huge spontaneous polarization $P^{\rm DFT}$ = 179 $\mu$C/cm$^{2}$.\cite{uratani.bicoo3.pcalc} 
The polar distortion can persist up to 733 K, when the sample finally decomposes. 
Since such a large polarization cannot be reversed by external electric fields, \bco\ should be regarded as a pyroelectric rather than ferroelectric system. 
\bco\ also exhibits C-type antiferromagnetic (C-AFM) ordering below the N$\rm\acute{e}$el temperature $T_{\rm N}$=420K, with ferromagnetic chains parallel to the $c$ axis that are antiferromagnetically coupled, as shown in Fig. \ref{fig:crys}.  
Therefore, BiCoO$_{3}$ is a good candidate material for spintronic applications, owing to the room-temperature-operating multifunctional properties.


%
\section{\label{sec:method}Method}

Density-functional-theory (DFT) calculations were performed using the VASP code \cite{vasp} within the  generalised gradient approximation (GGA). 
The rotationally invariant GGA$+U$ method was employed to account for correlation effects.\cite{dudarev} 
The Coulomb parameter $U$ for Co-$3d$ orbital states was fixed to 5.0 eV, 
as commonly used in the calculations of transition-metal oxides. 
We checked that other $U$ values do not significantly change our results, as discussed afterward. 
Lattice parameters are fixed to experimental values, $a=b=$ 3.7310   
\AA\ and $c=$ 4.7247 \AA.\cite{oka2010} A $\sqrt{2}\times\sqrt{2}\times1$ super cell has been used to host the C-AFM magnetic configuration of Co spins.
The internal atomic coordinates were fully optimized under C-AFM configuration without SOC, until forces acting on atoms were lower than 1$\times$10$^{-3}$eV/\AA.

\section{\label{symmetry} Symmetry analysis}
The space group of nonmagnetic \bco\ is $P4mm$, which has eight symmetry operations: 
$\{E, C_{2z}, C_{4z\pm}, \sigma_{x, y},  \sigma_{d, d'}$\}. 
The magnetic space group of C-AFM is $P4'mm'$ \footnote{The magnetic space group $P4'mm'$ is defined with the number \#99.12.834 in the OG setting. Note that the super cell in Fig.\ref{fig:crys} has a (1/2, 0, 0) translation with respect to the standard setting.}, 
which has  the sixteen operations: 
 $\{E, C_{2z}, C_{4z\pm}+t, \sigma_{x, y}+t, \sigma_{d, d'}, 
 t\cdot\theta, (C_{2z}+t)\cdot\theta, C_{4z\pm}\cdot\theta, \sigma_{x, y}\cdot\theta,  (\sigma_{d, d'}+t)\cdot\theta$\} 
where $t$ denotes a translation $t$ = (1/2, 1/2, 0) in a doubled magnetic unit cell and  
$\theta$ denotes the antiunitary time-reversal operator. 
The $C_{2z}$ and $\sigma_{x, y}$ symmetries relate spin sites belonging to the same magnetic sublattice, whereas $C_{4z\pm}$ and $\sigma_{d, d'}$ relate spin sites belonging to different magnetic sublattices in the C-AFM configuration. Since the latter displays two magnetic sublattices with opposite magnetization and given the transformation properties of an axial vector under a mirror symmetry, both $\sigma_{x, y}$ and $C_{4z\pm}$ operations act as spin-flip symmetry operations, forming a symmetry element of the system only when combined with the time-reversal operation.
 The magnetic Brillouin zone will therefore be contained in the nonmagnetic one, with coinciding $\Gamma$, Z high-symmetry points and with the M', A' points in the magnetic Brillouin zone coinciding with the X, R points of the nonmagnetic one. Since the symmetry operations of the nonmagnetic super cell are exactly those of the primitive unit cell, the symmetry properties of the eigenfunctions at these special high-symmetry points can be immediately determined by finding the irreducible representations of the little point groups in the nonmagnetic unit cell. In particular, the introduction of the magnetic sublattice will not induce any additional splitting of the energy bands at these special points, therefore preserving at least the two-fold degeneracy of the nonmagnetic system\cite{Dimmock1962}. This prerequisite allows for possible SOC-induced spin-splitting effects analogous to those realized in non-magnetic system. 

As a practical example, we construct an effective Hamiltonian, including SOC, modeling the band structure around $\Gamma$ point for a basis function $\vert\phi\rangle$ transforming as $z$ or $3z^2-r^2$ and belonging to the one-dimensional $\Gamma_1$ class. The analysis holds true also for Z, A and M points, the associated little group being always $C_{4v}$ for these high-symmetry $\bm k$ points. Since the two-fold Kramers degeneracy of the nonmagnetic system is kept at these points, we are allowed to consider as a basis $\vert \phi\rangle \otimes \vert\uparrow,\downarrow\rangle$, where the opposite spin states may be considered as belonging to the two different magnetic sublattices. The little group of $\Gamma$ in the magnetic phase contains four unitary operations $\{E, C_{2z}, \sigma_{d}, \sigma_{d'}\}$ and four antiunitary operations $\{C_{4z+}\cdot\theta, C_{4z-}\cdot\theta, \sigma_{x}\cdot\theta, \sigma_{y}\cdot\theta\}$. 
We notice that time-reversal by itself is not a symmetry operation of the magnetic system, and it enters as a symmetry element only in combination with point-group elements of the nonmagnetic space group that would induce a spin flipping. Taking into account the transformation rules relevant for the magnetic point group at $\Gamma$ and listed in Table \ref{tab:trules}, one immediately realizes that the symmetry-allowed linear spin-momentum coupling for bands belonging to the $\Gamma_1$ symmetry class acquires the typical linear-Rashba expression:
\begin{eqnarray}\label{eq:model}
H_{\Gamma} &=& E_0+\alpha\,(s_x k_y- s_y k_x)
\end{eqnarray}
where $\bm s$ are Pauli matrices accounting for spin-1/2 degrees of freedom, $\alpha$ denotes the Rashba spin-momentum coupling constant and $E_0=(\hbar^2/m_{\perp}^\ast)(k_x^2+k_y^2)+(\hbar^2/m_{z}^\ast)k_z^2$ is the free-electron like (parabolic) contribution with effective masses $m_\perp^\ast, m_z^\ast$. A similar analysis can be made for the other irreducible representations of the wavefunctions at the special high-symmetry points preserving the same degeneracy of the nonmagnetic system. On the basis of the symmetry analysis, the antiferromagnet polar \bco~ represents a good candidate for exploring the bulk Rashba effect in polar magnetic transition-metal oxides, such as multiferroic materials. 

\begin{table}[b]
\caption{Transformation rules for crystal momentum $\boldsymbol{k}$ and spin-$1/2$ operators under the considered magnetic point-group symmetry operations.
\color{black}
}\label{tab:trules}
\begin{tabular}{lp{0.5cm}cp{0.3cm}c}
\hline
&&$\{k_x,k_y,k_z\}$ &&$\{s_x,s_y,s_z\}$ \\
\hline
$E$ & & $\{k_x,k_y,k_z\}$ &&$\{s_x,s_y,s_z\}$ \\
$C_2$ & &$\{-k_x,-k_y,k_z\}$ && $\{-s_x,-s_y,s_z\}$\\
$\sigma_{d}$ && $\{ -k_y,-k_x,k_z \}$ && $\{s_y,s_x,-s_z\}$\\
$\sigma_{d'}$ && $\{ k_y,k_x,k_z \}$ && $\{-s_y,-s_x,-s_z\}$\\
\hline
 $C_{4z+}\cdot\theta$ & &$\{-k_y,k_x,-k_z\}$& &$\{-s_y,s_x,-s_z\}$\\
 $C_{4z-}\cdot\theta$ & &$\{k_y,-k_x,-k_z\}$& &$\{s_y,-s_x,-s_z\}$\\
 %
 $\sigma_x\cdot\theta$&& $\{k_x,-k_y,-k_z\}$ &&$\{-s_x,s_y,s_z\}$\\ 
 $\sigma_y\cdot\theta$ && $\{-k_x,k_y,-k_z\}$ && $\{s_x,-s_y,s_z\}$\\
  \hline
\end{tabular}
\end{table}

\section{\label{sec:electrronic} Electronic States and Rashba Spin Splitting}

\begin{figure}[ht]
\begin{center}
{
\includegraphics[width=60mm, angle=0]{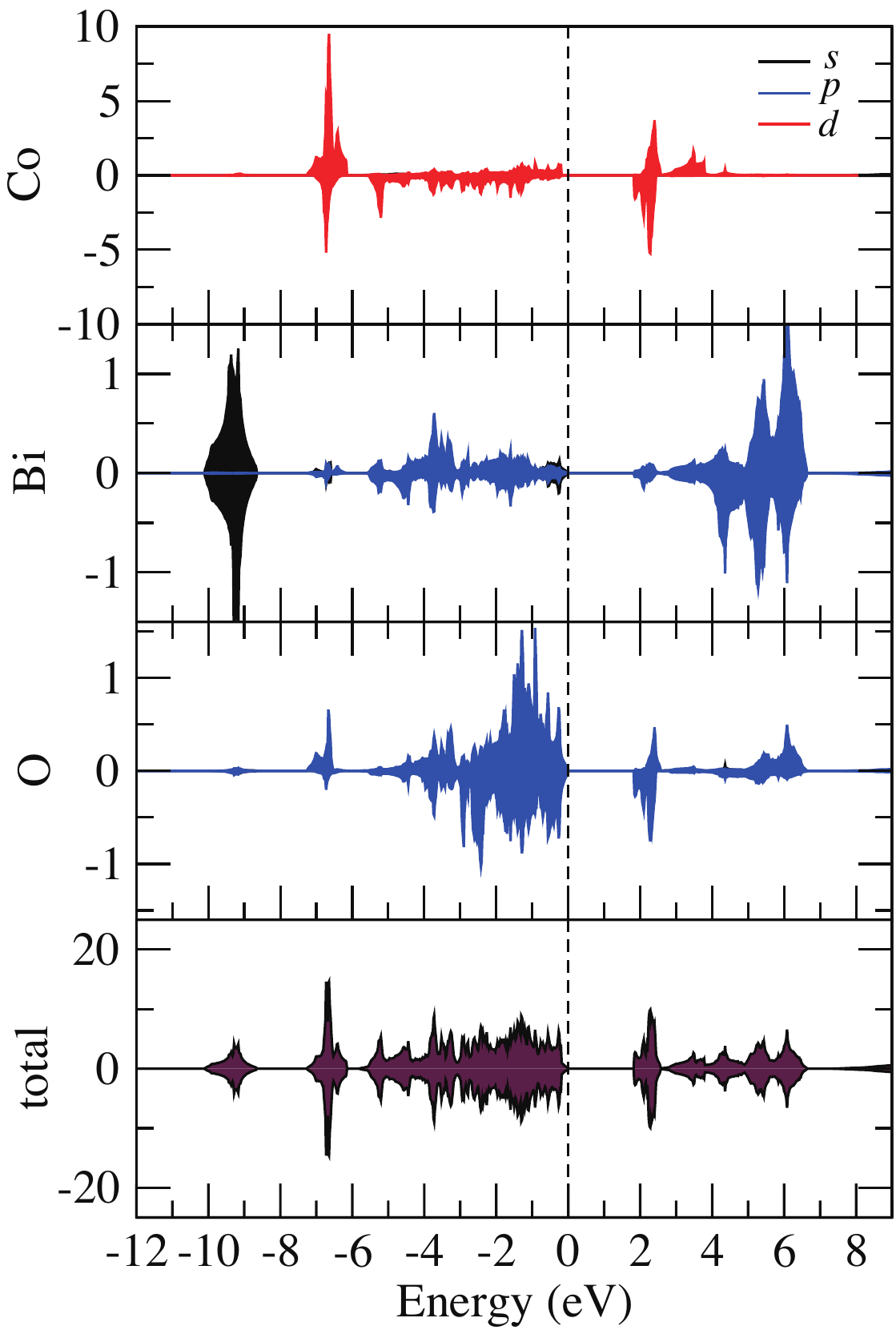}
}
\caption{\label{fig:dos}
Projected (onto each atomic orbital state) and total density of states.  
}
\end{center}
\end{figure} 
In this Section, we analyze in details the electronic structure of polar C-AFM BiCoO$_{3}$. The calculated density of states is shown in
Figure \ref{fig:dos}.
Near the Fermi energy, it is evident that Co-3$d$, Bi-6$p$ and O-2$p$ states are well hybridized.  
The strong hybridization between Bi-$6p$ and O-$2p$ is the driving force of the polar ionic distortion, supported by the instability of Bi-6$s$ lone pair. 
High-spin Co$^{3+}$ ($d^{6}$) ion has partially occupied $t_{2g}^{\downarrow}$ orbital state, which is split into occupied $\left| xy\right>$ and unoccupied $\left| yz\right>$,  $\left| zx\right>$ states by Jahn-Teller effect in the O$_{5}$ pyramid. 

\begin{figure}[ht]

{
\begin{center}
\includegraphics[width=90mm, angle=0]{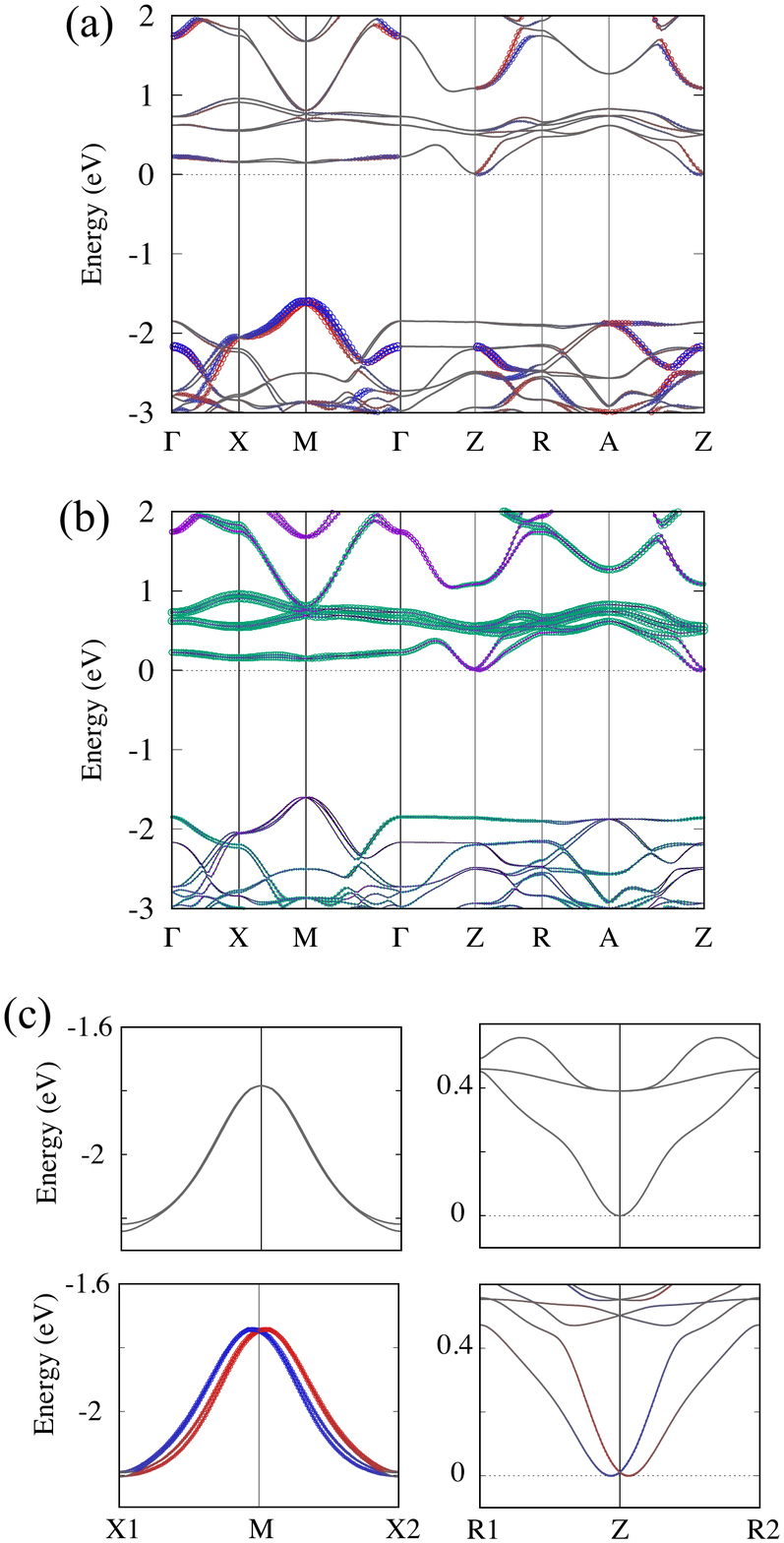}
\vspace {0.0cm}
\end{center}
}
\caption{\label{fig:band}
 (a) Bandstructure with a projection to in-plane spin polarization ($S_{x}$, $S_{x}$) perpendicular to the $k$ vector, where blue (red) color indicates up (down) spin. 
 (b)  Bandstructure with a projection to orbital states, where green (purple) color indicates Co-$d$ (Bi-$p$) orbital state. 
 The origin of the energy scale is set at the conduction band minimum. 
 (c) Bandstructure at VBM (left) and CBM (right) without SOC (top) and with SOC (bottom). 
}
\end{figure} 

Figure \ref{fig:band} shows the calculated bandstructure. 
Spin-splitting can be seen both at conduction-band minimum (two-fold degenerate) at the Z point and at valence-band maximum  (four-fold degenerate) at the M point, consistently with our previous symmetry analysis. 
The former band splitting originates from the strong dispersive Bi-$p_{z}$ component mixed with a rather flat Co-$d$:${3z^{2}-r^{2}}$ band, thus corresponding exactly to the case discussed in Section \ref{symmetry} and being accurately described by the Rashba model Eq. (\ref{eq:model}). Additionally, the free-electron like parabolic band character and strong SOC of Bi-$p$ band are expected to induce a sizable Rashba spin-splitting, as discussed in the next section. 
We checked the $U$ dependence of the conduction band character and found that 
the Co-$d$:${3z^{2}-r^{2}}$ orbital state dominates the conduction-band minimum in a range of $0 \leqslant U \leqslant 2$ eV, whereas
the Bi-$p$ orbital state appears at the conduction-band bottom in a range of $3\leqslant U \leqslant 7$ eV. 
Since the latter $U$ range seems more adequate for Co$^{3+}$ 
as seen in previous literature studies \cite{iniguez.bicoo3.prl2011, olsson.smcoo3.2016, ritzmann.lacoo3}, 
hereinafter we fix the $U$ value as 5.0 eV and discuss the Rashba splitting of conduction band minimum. 

On the other hand, the valence top band consists of $(p_{x}, p_{y})$  states of the apical oxygens with small mixing with Bi-$(s, p_{z})$ state. 
However, since the main topic of this paper is the interplay between Rashba effects and magnetism,
we focus on the simpler conduction-band splitting, 
 where the spin-splitting is much larger than that of valence top band. 


\section{Rashba spin splitting} 
\begin{figure}[ht]
\begin{center}
{
\includegraphics[width=70mm, angle=0]{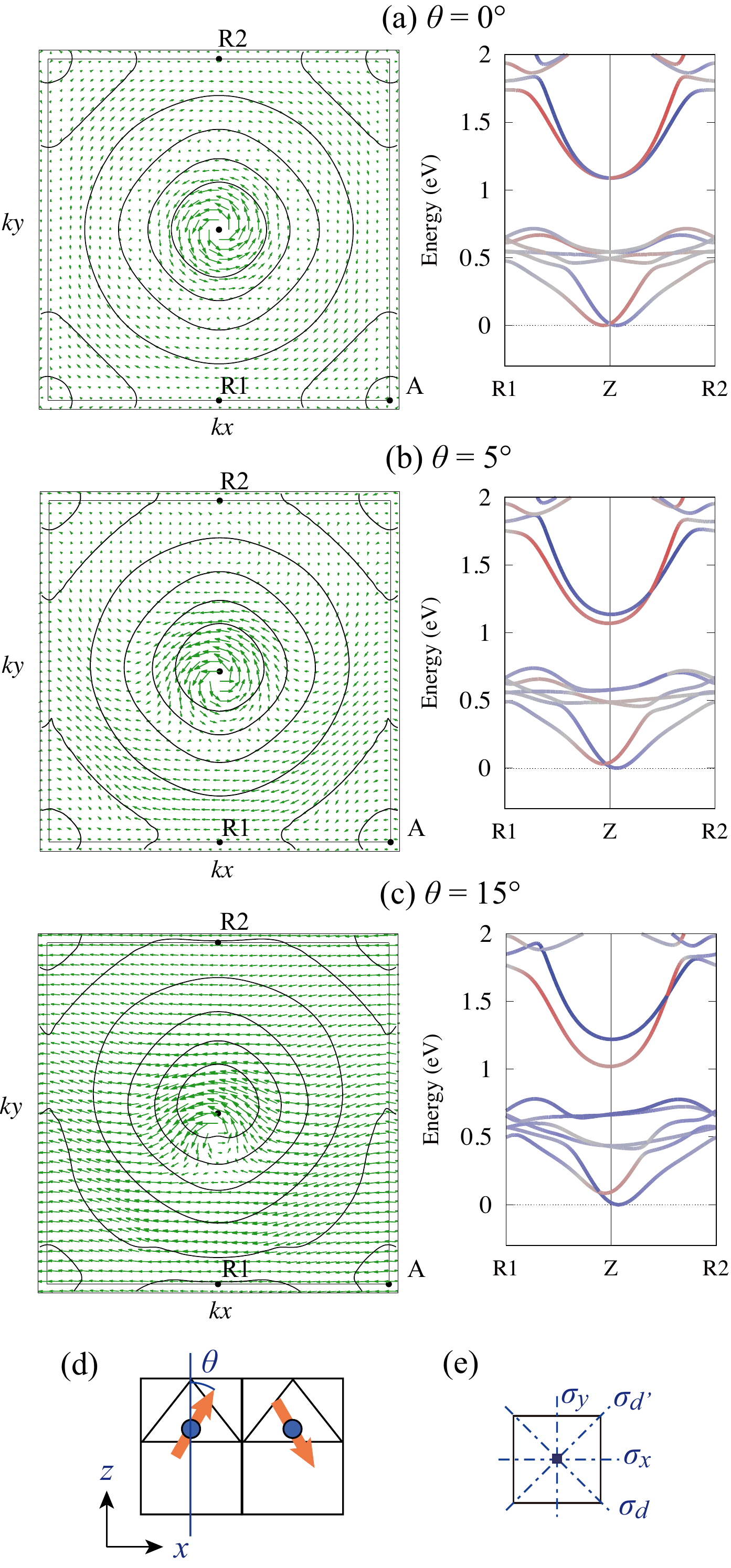}
}
\caption{\label{fig:spntxtcant}
Spin texture (shown by green arrows, left column) and energy contour plot (with respect to the conduction band bottom at 0.1 eV interval) of the CBM band 
in the $k_{z}=1/2$ plane (centered at the Z point) and spin polarized band structure (right column) as canting Co spins from [001] towards [100] direction with $\theta$=0, 5 and 15$^{\circ}$, shown in (a), (b), (c), respectively.  
Blue (red) shows unit of $k$ vector and energy is $\rm \AA ^{-1}$ and eV, respectively. 
(d) Canted spin AFM configuration with angle $\theta$. (e) Four mirror symmetries around the Z point in Brillouin zone.  
}
\vspace {0cm}
\end{center}
\end{figure} 

Figure \ref{fig:spntxtcant} (a) shows the spin texture of CBM. 
In agreement with our symmetry analysis, the characteristic ``vortex-like'' spin-texture of the Rashba effect develops around the Z point.
We estimated the Rashba parameter $\alpha = 2E_{\rm R} / K_{\rm R}$ along the Z-R line as $\alpha$=0.59 eV$\rm\AA$, where $K_{\rm R}=m_\perp^\ast\alpha/\hbar^2$ quantifies the mutual shift of the split bands and $E_{\rm R}=\alpha^2 m_\perp^\ast/(2\hbar^2)$ is the energy minimum of the split band.
This value is slightly smaller than that of BiAlO$_{3}$ with a predicted value, $\alpha$=0.74~eV$\rm\AA$, reported in Ref.\onlinecite{luiz.bialo3}.
However, we remark that BiAlO$_{3}$ is stabilized in BaTiO$_{3}$-like structure with $c/a$ ratio equal to 1.03 in the previous study, but it is found that the energy of BiAlO$_{3}$ is lower by 27 meV in a \bco-like structure 
i.e. with a much larger $c/a$ ratio of 1.27.
 It therefore seems that there are two metastable phases, with smaller and larger $c/a$ ratio, the most energetically stable being the \bco-like structure. 
In the structure with larger 
 c/a ratio, the Rashba coefficient of BiAlO$_{3}$ is very small ($\alpha$=0.12 eV$\rm\AA$), 
 the reduction in $\alpha$ being likely
caused by the change of the local environment around the Bi ion. 

The magnetic symmetry can be reduced when the Co spins are canted with respect to the C-AFM configuration in the real space. 
We assume the application of external magnetic field along the [100] direction,  
by canting
 two Co spins simultaneously, as illustrated in Fig. \ref{fig:spntxtcant} (d). 
This breaks 
the $C_{2z}$, $C_{4z\pm}\cdot \theta$ rotation, $\sigma_{x}\cdot \theta$ and $\sigma_{d, d'}$ mirror symmetries, 
but keeps the $\sigma_{y}\cdot \theta$ mirror symmetry. 
The effect of a magnetic field $B_x$ can be included in the minimal model Eq. (\ref{eq:model}) as $H_{Z_1}(B_x)=E_0+\alpha\,(s_x k_y- s_y k_x)+B_x\,s_x$, with eigenvalues $E_\pm=E0\pm\sqrt{\alpha^2 k_x^2+(\alpha k_y+B_x)^2}$.
As a result, we observe a distortion of the energy isocontour along the $k_y$ axis and a downward shift of the center of spin vortex, characteristic of Rashba-like spin-splitting (cfr. Fig. \ref{fig:spntxtcant}(a), (b), (c), respectively obtained for increasing spin-canting angles). 
From the energetic point of view, we note that, by
 comparing the total energy, it costs 1.6 meV/Co and 11.5 meV/Co for a spin canting with $\theta=5^{\circ}$ and $\theta=15^{\circ}$, respectively. 
Analogously to this case, 
 an in-plane magnetic field would distort the spin texture with a distortion of the energy contour and a shift of the spin-vortex center along a direction perpendicular to the applied field.

\section{Conclusion} 
We have demonstrated, by means of first-principles simulations complemented by symmetry arguments and model Hamiltonian, that bulk Rashba effect can occur in multifunctional materials, such as multiferroic BiCoO$_{3}$, where ferro-(pyro-)electric order and antiferromagnetic order coexist. While the antiferromagnetic symmetry doesn't lead to a finite spin polarization in the momentum space per se, a strong SOC on the Bi-$p$ bands leads to a Rashba-like $\bm k$-dependent spin splitting, where the spin polarization is perpendicular to the electric polarization direction. Furthermore, in an antiferromagnetic system, the Rashba spin texture can be modulated by applying an external magnetic field: the characteristic Rashba-like spin-vortex, centered on the high-symmetry point, can be shifted along a direction perpendicular to the applied magnetic field, in addition to the conventional spin canting. While the theoretical predictions in the present study call for experimental confirmation,  we speculate that this phenomenon can be rather common among the antiferromagnetic polar systems; in future works, one can explore the possibility of Rashba effects in various multiferroic systems, such as BiFeO$_{3}$ and PbVO$_{3}$.

\acknowledgments
This work was supported by JSPS Kakenhi (No. 17H02916 and 18H04227). 
A part of the computation in this work has been done by using the facilities of the Supercomputer Center, the Institute for Solid State Physics, the University of Tokyo.
KY acknowledges T. Oguchi for the fruitful discussions on Rashba effect. 
The crystallographic figure was generated using VESTA program.\cite{vesta}
 
\bibliography{biblio}
\end{document}